\documentclass[runningheads, orivec]{llncs}
\usepackage{esvect}
\usepackage[T1]{fontenc}
\usepackage[utf8]{inputenc}
\usepackage{booktabs}
\usepackage{multirow}
\usepackage{array}
\usepackage{tabularx}
\usepackage{makecell}
\usepackage{xcolor}
\usepackage{colortbl}
\usepackage{amsmath}
\usepackage{amssymb}
\usepackage{graphicx}
\usepackage{hyperref}
\usepackage{tikz}
\usepackage{enumitem}
\usepackage{float}
\usepackage{orcidlink}
\usepackage{todonotes}
\usepackage{paralist}
\usepackage{subfigure}
\usetikzlibrary{shapes.geometric,arrows.meta,positioning,fit,backgrounds,calc}

\definecolor{hdr}{RGB}{25,55,95}
\definecolor{rowA}{RGB}{242,246,252}
\definecolor{rowB}{RGB}{255,255,255}
\definecolor{cellEmpty}{RGB}{240,200,200}
\definecolor{cellSparse}{RGB}{255,235,180}
\definecolor{cellDense}{RGB}{200,230,200}
\definecolor{cellTarget}{RGB}{70,130,200}
\definecolor{cellTargetFuture}{RGB}{140,185,230}

\newcolumntype{C}[1]{>{\centering\arraybackslash}p{#1}}
\newcolumntype{L}[1]{>{\raggedright\arraybackslash}p{#1}}

\hypersetup{colorlinks=true,linkcolor=hdr,citecolor=hdr,urlcolor=hdr}

\newcommand{\tim}{\>\!$\times$\>\!}

\begin{document}

\title{Smart Adaptive Computing Across the Continuum:
LLMs in IoT-Edge-Cloud Resource Management}
\titlerunning{Smart Adaptive Computing Across the Continuum}

\author{Antonino Vaccarella\inst{1,2}~\orcidlink{0009-0004-0428-540X} \and
Lanpei Li\inst{1,2}~\orcidlink{0009-0005-4370-1020} \and
Vincenzo Lomonaco\inst{3}~\orcidlink{0000-0001-8308-6599} \and
Massimo Coppola\inst{1}~\orcidlink{0000-0002-7937-4157}}
\authorrunning{A. Vaccarella et al.}

\institute{
Institute of Information Science and Technologies ``Alessandro Faedo'' (ISTI), National Research Council of Italy (CNR), 56124 Pisa, Italy\\
\and
Department of Computer Science, University of Pisa, 56127 Pisa, Italy\\
\and
Department of AI, Data and Decision Sciences,\\ LUISS University, 00197 Rome, Italy\\
\email{antoninovaccarella@cnr.it, vlomonaco@luiss.it\\
\{lanpei.li, massimo.coppola\}@isti.cnr.it}}

\maketitle

\begin{abstract}
Managing resources across IoT, edge, and cloud layers calls for
continuous, context-aware decisions under constraints that rarely stay
fixed. Deep reinforcement learning (DRL) handles this class of problems
well, and large language models (LLMs) are increasingly used to augment
DRL pipelines, yet the architectural relationship between the two is
seldom made explicit. We build on Wang et al.'s taxonomy~\cite{wang2026drl} of Continuum Orchestration Systems employing DRL techniques and extend it with two further dimensions. The \emph{AI Augmentation Paradigm} measures how LLMs are exploited, while
the \emph{Feedback channel} captures whether and through which system path the execution feedback returns to the LLM in order to close the MAPE control loop~\cite{kephart2003vision} at the LLM Orchestration layer. We apply this taxonomy to six recent system architectures and find a common gap, as none combines full LLM orchestration with full agent-layer feedback in a Cloud Continuum setting. 
We relate this gap to
a missing cross-tier feedback abstraction, 
bridging the incommensurable per-tier signals and the LLM Orchestrator. 


\end{abstract}
\renewcommand{\thefootnote}{\roman{footnote}}
\footnotetext[0]{
This version of the contribution has been accepted for publication, after peer review but is not the Version of Record and does not reflect post-acceptance improvements, or any corrections. The work was presented at the 6th workshop on Flexible Resource and Application Management on the Edge (FRAME) 2026, co-located with the 32nd International European Conference on Parallel and Distributed Computing -- Euro-Par 2026. The Version of Record will appear in the workshop proceedings volume(s) of Euro-Par 2026. Use of this Accepted Version is subject to the publisher's Accepted Manuscript terms of use \url{https://www.springernature.com/gp/open-research/policies/accepted-manuscript-terms}.
}
\renewcommand{\thefootnote}{\arabic{footnote}}
\setcounter{footnote}{0}%

\section{Introduction}
\label{sec:intro}
Modern services no longer run only in homogeneous cloud data centers; operating them across a Cloud Continuum that spans IoT gateways, edge servers, and cloud instances requires continuous decisions about placement, scheduling, scaling, and traffic routing under constraints that shift with load patterns, user mobility, network conditions, and hardware availability. Resource management in the Cloud Continuum is therefore a multi-objective control problem: the system must balance performance and efficiency with policy-level requirements such as user intent, locality rules, and security constraints.

Deep reinforcement learning (DRL, or just RL in our context) is well suited to this setting for two reasons. First, it is designed to optimise sequences of decisions over time, which matches the inherently temporal nature of resource management: placement, scheduling, and routing choices made at one instant affect the state of the system at the next. Second, it requires no \textit{a priori} model of system dynamics, meaning the agent learns directly from interaction with the environment rather than from a hand-crafted mathematical model of how resources behave --- an important advantage when the Cloud Continuum is too complex or dynamic to model explicitly. Wang et al.~\cite{wang2026drl} document this trend across more than 100 papers through a two-dimensional taxonomy that separates control scope (SARL vs.\ MARL) from training paradigm (standard vs.\ federated).
Existing work shows that DRL and MARL are strong tools for fast, local, and adaptive control. However, they do not solve the full orchestration problem in next-generation Cloud Continuum systems. Their policies are often hard for operators to understand, depend heavily on the reward functions and state representations used during training, and struggle with constraints expressed as natural-language intents, compliance rules, or evolving security requirements. 
This suggests that DRL should be supported by a higher-level reasoning layer. LLM-based agentic systems can provide this layer by interpreting operator intents, reasoning over policy and context, explaining decisions, calling tools, and coordinating specialised agents. In the Cloud Continuum, the role of LLMs is therefore not to replace fast DRL/MARL controllers, but to provide explainable, policy-aware, and adaptive orchestration above them.

Recent systems point toward this direction. \textit{IntentContinuum}~\cite{akbari2025intentcontinuum} uses an LLM for intent-driven resource management across the compute continuum; works such as \cite{habib2025llmhrl}, SALLMA~\cite{becattini2025sallma} and AgentEdge~\cite{gort2025agentedge} study LLM-based multi-agent orchestration.
There is a shift from isolated LLM decisions to agentic systems that use delegation, tools, and orchestration feedback. 
These works also raise important design questions: whether
LLMs act as direct decision-makers, orchestrators delegating to downstream
agents, or closed-loop orchestrators, and whether their integration with
DRL results in a genuine feedback loop or merely a one-shot handoff. How do these choices affect resource management across Cloud Continuum?
Wang et al.~\cite{wang2026drl}
observe that large language models are beginning to appear alongside DRL in network optimisation, but the architectural relationship between the two remains underspecified. 

Our contribution in this paper is to address that gap by extending their taxonomy with two further dimensions. The \emph{AI Augmentation Paradigm} captures whether the LLM delegates execution to autonomous downstream agents, yielding three levels (LLM0--2) from pure DRL to LLM orchestrator systems. The \emph{F} dimension captures the kind of feedback channel through which execution data returns to the LLM. We apply the extended taxonomy to six recent works to identify existing architectural patterns and highlight gaps in the literature.

The works we analyzed so far use the LLM in different roles within the system architecture. From both a system and a functional viewpoint, an LLM processing human input is quite different from a true LLM orchestrator. The kind of control loop that is designed within the system conditions the actual capability of the LLM as a true orchestrator, thus the \emph{F} dimension is truly relevant in classifying current and future systems. 
In Sect.~\ref{sec:cloudorchestration} we briefly summarize the issues of dynamic resource orchestration for Clouds and Continuum platforms.  Sect.~\ref{sec:dim3} discusses the abstract architecture and the classification of LLM-enabled Orchestration solutions.
Sect.~\ref{sec:papers} analyzes the surveyed works with respect to the proposed taxonomy. Sect.~\ref{sec:gap} and \ref{sec:design-space} respectively
focus on what are today's gaps and issues in designing such systems, and identify a key need for a mediator functionality bridging the LLM/agents gap.
Sect.~\ref{sec:conclusion} summarizes our conclusions and future work plans.

\section{Cloud Continuum Orchestration}\label{sec:cloudorchestration}
The paradigm of Cloud computing, where all types of computing resources and services are automatically located, negotiated, acquired and orchestrated on-demand, has been evolving from the ``simple'' management of data-center based services toward including a family of geographically widespread computing devices that help close the proximity gap toward the users and the data sources.  

Continuum computing platforms are the result of including such a range of devices: data-center based ones, local cloud-like resources, vehicle-based, IoT, and mobile devices, with recent research tapping into esoteric devices like satellite-based ones.
The typical constraints of the Cloud orchestration problem are made more complex by the Continuum conditions: we need to manage a multilayer distributed platform, taking into account resource heterogeneity both across the platform layers and within the layers. Multi-tenancy as well as data privacy and locality constraints are pervasive. Applications exhibit uneven dynamic workload and dynamic user behaviour, prompting for autonomic-style, real-time reaction and planning as opposed to steady state system modeling. 

Orchestration on Continuum platforms implies a dynamic, multi-factor optimization problem with complex, sometimes unstructured constraints. Examples of such constraints are the interpretation of user intent, and the need to cope with security constraints derived from classified vulnerabilities (CVEs) that have to be combined with the  configuration of computing, networking and storage devices in order to mitigate vulnerabilities and shield from supply-chain attacks.

Several research approaches have been pursued to deal with complex, multifactor optimization and orchestration, including linear and non-linear optimization based ones, model based ones and heuristic approaches.
Recent works have addressed the need to react quickly and adapt to complex, unanticipated behavior by resorting to Reinforcement Learning (RL)-augmented agents.

Wang et al.~\cite{wang2026drl} survey a large corpus of works exploiting RL agents for Continuum Orchestration, covering both Single and Multiple RL Agents approaches (SARL vs MARL). 
However, it can be argued that some important features are still impractical or beyond the reach for RL  systems, notably as intricate security constraints  (those trying to prevent CVE exploitation and supply-chain attacks) and natural language user interaction are sought for. LLMs seem to be able to address these limitations, as \cite{wang2026drl} already identifies their use as a next research direction for Continuum Orchestration. 
They organise their taxonomy of DRL-based Cloud Continuum resource management systems along two dimensions:

\textbf{%
Control Scope} separates Single-Agent Reinforcement Learning (SARL) from Multi-Agent RL (MARL). SARL uses one global policy, such as DQN, PPO, A3C. MARL uses multiple agents that are either trained independently or employ the same model\footnote{For example, in Centralised Training with Decentralised Execution (CTDE) methods, agents share global
information during training to coordinate learning but act solely on their own
local observations once deployed. Representative CTDE methods include VDN \cite{sunehag2018vdn}, QMIX \cite{rashid2018qmix}, and MADDPG \cite{lowe2017maddpg}.%
}.

\textbf{%
Training Paradigm} separates standard training, where data can be shared and is possibly centralized for the training, from federated training, where learning is coordinated through model update methods such as Federated Averaging (FedAvg), or gossip techniques, without ever centralising or sharing the raw data.

This bidimensional taxonomy captures how DRL controllers are organized and trained. It does not, however, describe the architectural role of an LLM when language model components are added to the control loop.

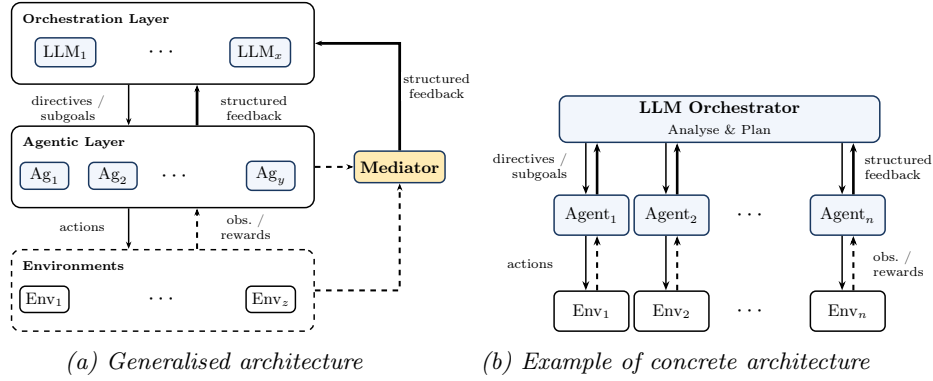
\begin{figure}[t]
\centering
\begin{minipage}[b]{0.5\textwidth}
\centering
\resizebox{\linewidth}{!}{%
\begin{tikzpicture}[
  font=\footnotesize,
  llmbox/.style={rectangle, rounded corners=3pt, minimum width=1.15cm,
                 minimum height=0.55cm, draw=hdr, thick, fill=rowA,
                 align=center, font=\footnotesize\bfseries},
  agbox/.style={rectangle, rounded corners=3pt, minimum width=0.92cm,
                minimum height=0.50cm, draw=hdr, thick, fill=rowA,
                align=center, font=\footnotesize\bfseries},
  envbox/.style={rectangle, rounded corners=3pt, minimum width=0.92cm,
                 minimum height=0.50cm, draw=black, thick, fill=white,
                 align=center, font=\footnotesize},
  ell/.style={font=\large\bfseries, text=black},
  layer/.style={draw=black, thick, rounded corners=5pt,
                minimum width=5.75cm, minimum height=1.55cm},
  envlayer/.style={draw=black, thick, dashed, rounded corners=5pt,
                   minimum width=5.75cm, minimum height=1.55cm},
  layerlabel/.style={font=\scriptsize\bfseries, anchor=north west,
                     inner sep=1.2pt},
  flowlabel/.style={font=\scriptsize, fill=white, inner sep=1.3pt,
                    align=center, text depth=0pt},
  arr/.style={-{Stealth[length=4pt]}, thick, color=black},
  fdbk/.style={-{Stealth[length=4pt]}, line width=1.5pt, color=black},
  obs/.style={-{Stealth[length=4pt]}, line width=1.1pt, dashed, color=black},
  medbox/.style={rectangle, rounded corners=3pt,
                 minimum width=1.3cm, minimum height=0.60cm,
                 draw=hdr, thick, fill=cellSparse,
                 align=center, font=\footnotesize\bfseries},
]

\node[llmbox] (llm1) at (1.15, 4.68) {$\mathrm{LLM}_1$};
\node[ell]           at (3.00, 4.68) {$\cdots$};
\node[llmbox] (llmx) at (4.85, 4.68) {$\mathrm{LLM}_x$};

\node[agbox]  (ag1)  at (0.75, 2.33) {$\mathrm{Ag}_1$};
\node[agbox]  (ag2)  at (2.05, 2.33) {$\mathrm{Ag}_2$};
\node[ell]           at (3.25, 2.33) {$\cdots$};
\node[agbox]  (agy)  at (5.05, 2.33) {$\mathrm{Ag}_y$};

\node[envbox] (ev1) at (0.75, -0.02) {$\mathrm{Env}_1$};
\node[ell]          at (3.00, -0.02) {$\cdots$};
\node[envbox] (evz) at (5.05, -0.02) {$\mathrm{Env}_z$};

\begin{scope}[on background layer]
  \node[layer]    (orchlayer) at (3.00,4.85) {};
  \node[layer]    (aglayer)   at (3.00,2.50) {};
  \node[envlayer] (envlayer)  at (3.00,0.15) {};
\end{scope}

\node[layerlabel] at ([xshift=5pt,yshift=-5pt]orchlayer.north west)
  {Orchestration Layer};
\node[layerlabel] at ([xshift=5pt,yshift=-5pt]aglayer.north west)
  {Agentic Layer};
\node[layerlabel] at ([xshift=5pt,yshift=-5pt]envlayer.north west)
  {Environments};

\coordinate (oaDownTop) at (2.35,0 |- orchlayer.south);
\coordinate (oaDownBot) at (2.35,0 |- aglayer.north);
\coordinate (oaUpBot)   at (3.65,0 |- aglayer.north);
\coordinate (oaUpTop)   at (3.65,0 |- orchlayer.south);
\draw[arr]  (oaDownTop) -- (oaDownBot);
\draw[fdbk] (oaUpBot) -- (oaUpTop);
\node[flowlabel, anchor=east] at (1.95,3.62)
  {directives /\\ subgoals};
\node[flowlabel, anchor=west] at (4.05,3.62)
  {structured\\ feedback};

\coordinate (aeDownTop) at (2.35,0 |- aglayer.south);
\coordinate (aeDownBot) at (2.35,0 |- envlayer.north);
\coordinate (aeUpBot)   at (3.65,0 |- envlayer.north);
\coordinate (aeUpTop)   at (3.65,0 |- aglayer.south);
\draw[arr] (aeDownTop) -- (aeDownBot);
\draw[obs] (aeUpBot) -- (aeUpTop);
\node[flowlabel, anchor=east] at (1.95,1.37)
  {actions};
\node[flowlabel, anchor=west] at (4.05,1.37)
  {obs.~/\\ rewards};

\node[medbox] (med) at (7.5, 2.50) {Mediator};
\draw[obs]  (aglayer.east)  -- (med.west);
\draw[obs]  (envlayer.east) -| (med.south);
\draw[fdbk] (med.north) |-
  node[flowlabel, pos=0.30, anchor=west]
    {structured\\ feedback}
  (orchlayer.east);

\end{tikzpicture}%
}
\end{minipage}%
\hfill
\begin{minipage}[b]{0.48\textwidth}
\centering
\resizebox{\linewidth}{!}{%
\begin{tikzpicture}[
  font=\footnotesize,
  llmbox/.style={rectangle, rounded corners=4pt, minimum width=6.0cm,
                 minimum height=0.85cm, draw=hdr, thick, fill=rowA,
                 align=center, font=\small\bfseries},
  agbox/.style={rectangle, rounded corners=3pt, minimum width=1.4cm,
                minimum height=0.75cm, draw=hdr, thick, fill=rowA,
                align=center, font=\footnotesize\bfseries},
  envbox/.style={rectangle, rounded corners=3pt, minimum width=1.4cm,
                 minimum height=0.75cm, draw=black, thick, fill=white,
                 align=center, font=\footnotesize},
  arr/.style={-{Stealth[length=5pt]}, thick, color=black},
  fdbk/.style={-{Stealth[length=5pt]}, line width=1.4pt, color=black},
  obs/.style={-{Stealth[length=5pt]}, line width=1.1pt, dashed, color=black},
  edgelabel/.style={font=\scriptsize, text=black, fill=white, inner sep=1pt},
]

\node[llmbox] (llm) at (3.1, 4.8)
  {LLM Orchestrator \\ {\mdseries\scriptsize Analyse \& Plan}};

\node[agbox] (ag1) at (0.7, 3.0) {$\mathrm{Agent}_1$};
\node[agbox] (ag2) at (2.2, 3.0) {$\mathrm{Agent}_2$};
\node[font=\large\bfseries]      at (3.7, 3.0) {$\cdots$};
\node[agbox] (agn) at (5.5, 3.0) {$\mathrm{Agent}_n$};

\node[envbox] (ev1) at (0.7, 1.2) {$\mathrm{Env}_1$};
\node[envbox] (ev2) at (2.2, 1.2) {$\mathrm{Env}_2$};
\node[font=\large\bfseries]       at (3.7, 1.2) {$\cdots$};
\node[envbox] (evx) at (5.5, 1.2) {$\mathrm{Env}_n$};

\draw[arr]  ([xshift=-3pt]ag1.north |- llm.south) -- ([xshift=-3pt]ag1.north);
\draw[arr]  ([xshift=-3pt]ag2.north |- llm.south) -- ([xshift=-3pt]ag2.north);
\draw[arr]  ([xshift=-3pt]agn.north |- llm.south) -- ([xshift=-3pt]agn.north);

\draw[fdbk] ([xshift=3pt]ag1.north) -- ([xshift=3pt]ag1.north |- llm.south);
\draw[fdbk] ([xshift=3pt]ag2.north) -- ([xshift=3pt]ag2.north |- llm.south);
\draw[fdbk] ([xshift=3pt]agn.north) -- ([xshift=3pt]agn.north |- llm.south);

\draw[arr]  ([xshift=-3pt]ag1.south) -- ([xshift=-3pt]ev1.north);
\draw[arr]  ([xshift=-3pt]ag2.south) -- ([xshift=-3pt]ev2.north);
\draw[arr]  ([xshift=-3pt]agn.south) -- ([xshift=-3pt]evx.north);

\draw[obs]  ([xshift=3pt]ev1.north) -- ([xshift=3pt]ag1.south);
\draw[obs]  ([xshift=3pt]ev2.north) -- ([xshift=3pt]ag2.south);
\draw[obs]  ([xshift=3pt]evx.north) -- ([xshift=3pt]agn.south);

\node[align=right, edgelabel] at (-0.45, 3.9)
  {directives /\\ subgoals};
\node[align=left, edgelabel] at (6.45, 3.9)
  {structured\\ feedback};
\node[align=right, edgelabel] at (-0.45, 2.1)
  {actions};
\node[align=left, edgelabel] at (6.45, 2.1)
  {obs.\ /\\ rewards};

\end{tikzpicture}%
}
\end{minipage}

\smallskip
{\footnotesize\textit{(a) Generalised architecture}}
\hfil\hfil
{\footnotesize\textit{(b) Example of concrete architecture}}
\caption{(a)~Generalised two-layer architecture ($x$ LLMs, $y$ agents, $z$ environments).
(b)~Example of architecture (LLM2 type) with one orchestrator, $n$ agents, and $n$ environments.%
}
\label{fig:closed-loop}
\end{figure}

\section{Generic architecture and AI Augmentation Paradigm}
\label{sec:dim3}
In a nutshell, the \textit{Orchestrator} in Cloud and Continuum platforms is the component solving the problem of devising a good allocation plan 
for one or more distributed applications over the set of resources within the platform, as well as initiating the actions required by the plan. As the needs of the applications and the availability of the devices will evolve over time, dynamic orchestration is typically needed in the Continuum \cite{ullah_orchestration_2023-1}.
Since our work is at the crossroad of AI and Distributed Computing, we shall underline that in the AI field an Orchestrator coordinates multiple AI entities (e.g. RL agents or distinct LLM agents) to accomplish a coordinated task.  
While both definitions may apply in some of the papers we discuss, for clarity we will always use the former one. 

Fig.~\ref{fig:closed-loop} illustrates a generic architecture for Continuum LLM-based orchestration.
We see the abstract form of such systems in (Figure~\ref{fig:closed-loop}a), where the layer of controlled execution systems is labeled \textit{Environments} and two  conceptual layers are defined above it: an \emph{Orchestration Layer} housing
one or more LLMs that analyse, plan, and issue directives or subgoals; and an
\emph{Agentic Layer} comprising one or more specialised agents that receive
those directives, act within their environments, and return structured
feedback upward. The cardinalities $x,y,z$ of the layers do not need to match with each other, and the actual interaction graph among the entities is not specified.

A concrete instantiation of the abstract architecture is shown in Fig.~\ref{fig:closed-loop}b, where the Orchestration layer is a single-LLM orchestrator, coordinating
$n$ agents in a 1-to-1 relationship with $n$ controlled environments. Here the centralized LLM issues subgoals while
agents interact with the environments and return observations and data as feedback.

\paragraph{Control loops towards the LLM} ---
A \emph{control system} is a \emph{closed-loop} one when the controller's decisions are updated from execution feedback, as opposed to being {open-loop control} where no feedback is assumed. In autonomic computing, the reference structure of control loops is the
Monitor--Analyse--Plan--Execute (MAPE) cycle~\cite{kephart2003vision}.
For LLM-augmented DRL systems, although an open-loop control style is possible, a closed control loop allows the system to refine its reactions, achieving dynamic adaptation \cite{yao_react_2023}.
The key questions become then what kind of feedback is there, how is it generated and from/to where in the system does the feedback travel. Our goal in this work is to analyze the overall SW architecture of Continuum orchestration, so we will not focus on the specific internal architecture of the LLM-augmented modules.

We focus instead on the control loops that bring information to the topmost layer of the system in Fig.~\ref{fig:closed-loop}a, providing feedback to an LLM-augmented controller that is (part of) the Continuum orchestrator to help refine its reactions. 
We distinguish orchestration control loops by the way they collect information, e.g., \begin{inparaenum}[(\bf i)] \item from the controlled systems, through adapter interfaces but without any processing, \item from the agent layer, letting the agents collect, filter and structure the feedback, \item from both the controlled systems and the agents, but exploiting dedicated agent(s)/service(s) for information filtering and processing.\end{inparaenum} 

\paragraph{Dimension F --- Feedback channel.}
We identify four choices (F0--F3) for the channel returning execution information
to the LLM, resting on two criteria: (a)~whether feedback from the real execution platform
reaches the LLM at all after agents have acted, and (b)~who mediates it.

\begin{description}[noitemsep,leftmargin=2em]
  \item[\textbf{F0 --- None.}]
    No feedback from the real environment reaches the LLM after execution.
    Internal mechanisms within the orchestrator --- critics, offline simulators,
    self-consistency checks --- operate on virtual or pre-execution data and are
    not considered environmental feedback.

  \item[\textbf{F1 --- Platform.}]
    Environment data reaches the LLM directly, or through a deterministic
    non-autonomous wrapper, from execution-platform APIs: Kubernetes event
    streams, SDN controller statistics, monitoring dashboards.
    No autonomous agent mediates this data before it arrives.

  \item[\textbf{F2 --- Sub-agent.}]
    Feedback is produced and routed by executing downstream sub-agents rather
    than drawn directly from platform APIs. The LLM receives a processed view
    of execution outcomes, filtered and structured by agents whose primary
    function is carrying out tasks, not reporting on them.

  \item[\textbf{F3 --- Dedicated mediator.}]
    A component dedicated exclusively to evaluation --- distinct from any
    executing agent --- receives raw platform and agent output and distils it
    into structured feedback for the orchestrator: reward signals, performance
    scores, alignment assessments.
\end{description}
Our viewpoint is that the adoption of multiple LLMs, actor-critic approaches, and simulation-based validation within the orchestrator does not form a closed-loop for control from this viewpoint, as these solutions employ virtual feedback from within the Orchestrator itself. Local feedback from execution to the RL agents is not considered either, unless it is explicitly forwarded to the LLM.

\paragraph{LLM integration} ---
The depth of LLM integration is captured by a single structural question: does the LLM
delegate execution to autonomous downstream agents? Three levels result.\par
\smallskip
\textbf{LLM0 — DRL Only.} No LLM component is present.
Covers all of \cite{wang2026drl}.

\textbf{LLM1 — Direct LLM Actor.} The LLM is the sole autonomous
decision-maker: it receives context, produces decisions, and enacts them
through APIs, tools, and components. No autonomous downstream
agents affect these goals.

\textbf{LLM2 — LLM Orchestrator.} The LLM delegates task execution to a
layer of autonomous downstream agents (RL-based, LLM-based, or hybrid). It
assigns objectives or subgoals but does not directly act on the environment:
agents execute independently within their assigned scope.
Whether and how execution outcomes reach back to the LLM is captured
by dimension~F (Fig.~\ref{fig:closed-loop}).

\section{Extending the Taxonomy towards Upcoming LLM Approaches}
\label{sec:papers}
Our extended taxonomy keeps Wang et al.'s \cite{wang2026drl} two axes \emph{Control Scope} and \emph{Training Paradigm} and adds two further ones, namely the \emph{AI Augmentation Paradigm} (LLM) and the \emph{Feedback Channel} (F) as defined in Sect.~\ref{sec:dim3}. Table~\ref{tab:grid} shows the resulting density map along the first three dimensions.

Control scope captures the multiplicity of autonomous decision-making entities, while the LLM level states their kind. At LLM0, these correspond to SARL and MARL in the original DRL taxonomy. At LLM1--LLM2, the controller may be anything (e.g., a LLM, a RL agent), hence the broader SA/MA labels.

This presentation is an architectural taxonomy, not a performance ranking. Its purpose is to show what combinations already exist and what parts of the design space remain underexplored. As we can see, LLM1--LLM2 extend into largely unpopulated territory; within this space, the LLM2-MA row concentrates the most significant structural gaps, which are examined in Section~\ref{sec:gap}.

\begin{table}[!b]
\centering
\caption{Density map of the Control Scope, Training Paradigm, and AI Augmentation Paradigm dimensions of the taxonomy.
  \colorbox{cellDense}{\phantom{X}}~dense cell;
  \colorbox{cellSparse}{\phantom{X}}~sparse;
  \colorbox{cellEmpty}{\phantom{X}}~empty%
  }
\label{tab:grid}
\footnotesize
\setlength{\tabcolsep}{3pt}
\renewcommand{\arraystretch}{1.3}
\begin{tabularx}{\columnwidth}{l >{\centering\arraybackslash}X >{\centering\arraybackslash}X >{\centering\arraybackslash}X >{\centering\arraybackslash}X}
\rowcolor{hdr}
\textcolor{white}{\textbf{}} &
\textcolor{white}{\textbf{Std-SA}} &
\textcolor{white}{\textbf{Std-MA}} &
\textcolor{white}{\textbf{Fed-SA}} &
\textcolor{white}{\textbf{Fed-MA}} \\
\\[-4.5mm]
\textbf{LLM0} &
  \cellcolor{cellDense}$>$50 works &
  \cellcolor{cellDense}$>$30 works &
  \cellcolor{cellDense}$\sim$12 works &
  \cellcolor{cellDense}$\sim$8 works \\
\textbf{LLM1} &
  \cellcolor{cellSparse}\cite{akbari2025intentcontinuum} &
  \cellcolor{cellSparse}\makecell{\cite{becattini2025sallma} \cite{gort2025agentedge}} &
  \cellcolor{cellEmpty}— &
  \cellcolor{cellEmpty}— \\
\textbf{LLM2} &
  \cellcolor{cellEmpty}— &
  \cellcolor{cellSparse}\cite{habib2025llmhrl} \cite{nourzad2025aura} \cite{peng2026cyberops} &
  \cellcolor{cellEmpty}\textit{N/A} &
  \cellcolor{cellEmpty}— \\
\bottomrule
\end{tabularx}
\end{table}

\begin{table}[!b]
\centering
\caption{Classification of the six works by LLM level and feedback dimension~F.
  Cells marked \textit{N/A} are structurally impossible (F2/F3 require
  downstream agents, so LLM2).}
\label{tab:llm-f}
\footnotesize
\setlength{\tabcolsep}{6.8pt}
\renewcommand{\arraystretch}{1.4}
\begin{tabularx}{\columnwidth}{l C{2.2cm} C{2.2cm} C{2.2cm} C{2.2cm}}
\rowcolor{hdr}
\textcolor{white}{\textbf{}} &
\textcolor{white}{\textbf{F0}} &
\textcolor{white}{\textbf{F1}} &
\textcolor{white}{\textbf{F2}} &
\textcolor{white}{\textbf{F3}} \\
\rowcolor{rowA}
\textbf{LLM1} &
  \cite{gort2025agentedge} &
  \cite{akbari2025intentcontinuum} \cite{becattini2025sallma} &
  \textit{N/A} &
  \textit{N/A} \\
\rowcolor{rowB}
\textbf{LLM2} &
  \cite{habib2025llmhrl} &
  — &
  \cite{peng2026cyberops} &
  \cite{nourzad2025aura} \\
\bottomrule
\end{tabularx}
\end{table}

None of the LLM-based works employ federated learning, so all recent papers we classify fall within the standard (non-federated) column of the taxonomy: the LLM is never trained, and only the three LLM2 systems train downstream RL agents. Three sit at LLM1, three at LLM2; of the latter, only two close the feedback loop at the orchestration level, as
we show in Table~\ref{tab:llm-f} about the interplay of the LLM level with the F dimension.
Each paper in the following is discussed also in relation to Fig.~\ref{fig:closed-loop}a, tracing what components occupy the Orchestration Layer, whether an autonomous Agentic Layer is present, and how --- or whether --- execution outcomes are fed back to the LLM.

\paragraph{LLM1\tim F0 : LLM without platform feedback.}

\textbf{AgentEdge}~\cite{gort2025agentedge} automates edge-cloud resource management through a sequential chain of four specialised LLM agents --- Intent, Observability, Planning, and Infrastructure Action --- each implemented as an independent model call whose output feeds the next stage. All four stages reside in the Orchestration Layer,
so the Agentic Layer is absent; the system is therefore \textbf{LLM1}, regardless of the number of model calls.
The Planning stage runs an ActSimCrit loop entirely on virtual state --- proposing, simulating on an offline infrastructure model, and critiquing until the plan passes an LLM-based quality check --- before the Infrastructure Action Agent executes on live infrastructure. At that point no signal travels back up the chain, leaving the Orchestration Layer without any post-execution information (\textbf{F0}).

\paragraph{LLM1\tim F1 : LLM with direct platform feedback. }
Despite targeting different problems, {IntentContinuum}~\cite{akbari2025intentcontinuum} and {SALLMA}~\cite{becattini2025sallma} share the same spot in our classification.
\textbf{IntentContinuum} monitors SLO compliance in a Kubernetes cluster. On each violation, Management and Orchestrator (MANO) polls three platform APIs --- Kubernetes, ONOS, and sFlow-RT --- and packages the result as a structured JSON payload for GPT-4o, which diagnoses the root cause and selects a corrective action. GPT-4o alone occupies the Orchestration Layer; MANO is a deterministic dispatcher, not an autonomous agent, so the Agentic Layer is empty (\textbf{LLM1}). Feedback reaches the model directly from platform APIs, with no autonomous mediator between environment and orchestrator (\textbf{F1}).
\textbf{SALLMA} is a general-purpose reference architecture for intent-driven AI workflows. An Intent Management Agent parses operator requests; a Workflow Management Agent decomposes them into a cognitive workflow executed inside Kubernetes-managed containers via LangChain, invoking specialised LLM instances with access to RAG stores and persistent memory. All components operate under the Workflow Management Agent's direct authority and pursue no independently assigned goals, so the entire pipeline sits in the Orchestration Layer (\textbf{LLM1}). Execution results are written to a passive SQL/NoSQL Knowledge Layer from which the Workflow Management Agent reads; a passive store does not mediate feedback autonomously, placing the channel at the same platform level as IntentContinuum (\textbf{F1}).

\paragraph{LLM2\tim F0 : LLM and Agents, no direct platform feedback.}

\textbf{Habib et al.}~\cite{habib2025llmhrl} target intent-based 5G Open Radio Access Network (RAN) automation. ALBERT \cite{Lan2019ALBERTAL} classifies operator intents via few-shot prompting; a Transformer predictor validates them against a traffic forecast; the validated goal is then handed off to a hierarchical DQN (h-DQN) that orchestrates five DRL RAN applications. ALBERT and the Transformer sit in the Orchestration Layer, while the h-DQN and its applications form the Agentic Layer, making this an \textbf{LLM2} system. The delegation, however, is a one-shot handoff followed by executing an HRL algorithm.
As agent-layer outcomes never return to the LLM, this system is an open-loop orchestrator  (\textbf{F0}).

\paragraph{LLM2\tim F2 : LLM with Agents providing platform feedback.}
\textbf{CyberOps-Bots}~\cite{peng2026cyberops} targets autonomous cyber-resilience in cloud networks. A Qwen3-8B orchestrator plans multi-step defence responses via ReAct \cite{yao_react_2023} and dispatches four types of heterogeneous RL agents --- Fortify, Recover, Purge, and Block --- deploying multiple instances dynamically across six network subnets (\textbf{LLM2}). Qwen3-8B is the single occupant of the Orchestration Layer; the four RL agents populate the Agentic Layer; the network regions are the Environments. After each action, a Perception module converts the updated network state to natural language, logs it in Short-Term Memory, and injects it into the LLM's context on the next planning cycle. The feedback is real and post-execution, but the only mediator is the executing agent tier itself,
which places the system as of \textbf{F2} type.

\paragraph{LLM2\tim F3 : LLM, Agents and Mediated Platform Feedback.}
\textbf{AURA}~\cite{nourzad2025aura} couples Claude Sonnet~4 with RL agents deployed at individual cellular base stations (\textbf{LLM2}). The LLM generates subgoals and reward-shaping parameters; agents accept or reject this guidance through a trust mechanism%
. Claude Sonnet and the Centralised Alignment Controller (CAC) sit in the Orchestration Layer; the RL agents at base stations form the Agentic Layer; the base stations are the Environments. After each cycle, agents report actions and outcomes to the CAC, which computes a delayed structured reward in $[-1,+1]$ --- assessing network efficiency, fairness, and adaptability --- and returns it to the LLM. What sets AURA apart from CyberOps-Bots is the role of the CAC: it never issues commands to base stations and is architecturally separate from the agents it assesses, its sole function being evaluation. That separation --- a dedicated mediator distinct from the executing tier --- is precisely what qualifies the channel as \textbf{F3}.
AURA operates within a single tier of cellular base stations, without multi-tier heterogeneity or Cloud Continuum scope.%

\section{Gap Analysis and Research Directions}
\label{sec:gap}

Based on the classification presented in the previous section, key gaps and some trends emerge in the limited amount of current literature:
\begin{asparaenum}[\bf (i)]
\item \textbf{Control loops towards the LLM are rare.}
Four out of the six classified works provide the LLM with no post-execution environmental feedback (AgentEdge, Habib et al.) or only with platform-level signals (IntentContinuum, SALLMA). Only AURA (F3) and CyberOps-Bots (F2) close the agent-layer feedback loop, both using the LLM as a full orchestrator (LLM2 class).
\item \textbf{Current LLM2 systems with agent-layer feedback do not address the Cloud Continuum.}
AURA and CyberOps-Bots demonstrate that this architectural pattern is viable, but both remain confined to a single deployment tier. The challenge of Continuum platforms remains unaddressed in many key aspects, like multi-tier compute heterogeneity, containerised workload placement, SLO management, resource tenancy, cost models, and SDN-based traffic routing. Structured post-execution feedback has emerged only in single-tier domains, where the feedback signal has a natural, well-defined form that
Cloud Continuum platforms do not provide.

\item \textbf{No work combines LLM orchestration (LLM2) with federated training of the agent layer}, but data privacy and locality improvements are already needed. 
Closing the feedback loop raises a data-privacy concern that promotes federated learning techniques.
While no current system yet classifies as LLM2\tim MA\tim Federated, 
two of the works look for improved locality exploitation.
\textbf{AgentEdge} identifies federated orchestration protocols as a primary future direction, and \textbf{CyberOps-Bots} explicitly names LLM centralization as its primary scalability concern. 
\end{asparaenum}

We expect that the combination of choices LLM2\tim MA\tim Federated will soon be explored, but
filling the voids in our classification will require addressing at least three engineering challenges still absent from the surveyed literature.

\begin{description}
\item[Latency mismatch.]
Directly deploying LLMs for fine-grained, real-time control remains infeasible due to their computational overhead and latency~\cite{nourzad2025aura}. Inserting the LLM on the critical path would stall the control loop; the gap must therefore be absorbed structurally --- for instance, by having the LLM operate asynchronously on episode summaries while quicker, local agents continue under the last issued subgoal or reward-shaping signal.

\item[Cost of LLM involvement.]
The cost here is monetary: each orchestration step is a billed API call. AURA, for instance, drives its orchestrator with Claude Sonnet~4~\cite{nourzad2025aura}, priced at \$3/\$15 per million input/output tokens
(see {\footnotesize\url{https://platform.claude.com/docs/en/about-claude/pricing}}).
At a per-step prompt of, say, 10{,}000 input and 1{,}000 output tokens, one call and one cycle cost about \$0.05, but 
the total cost scales with control frequency and the number of agents, easily reaching hundreds of dollars per day.
A threshold-triggered policy consults the LLM only when feedback deviates beyond a defined margin from the current directive, otherwise leaving the agents running under the last directive. Cutting the number of paid calls this way is a recognised approach to lowering LLM inference cost~\cite{chen2023frugalgpt}, and may allow extended open-loop stretches without significant performance loss.

\item[Feedback abstraction.]
Execution outcomes must be compressed into a prompt-sized representation that carries enough information to revise a directive \cite{NEURIPS2023_1b44b878}. In single-tier domains agents typically operate over a largely homogeneous KPI space; in the Cloud Continuum, by contrast, IoT, edge, and cloud agents tend to expose heterogeneous and often incommensurable metrics, and, to the best of our knowledge, a cross-tier feedback abstraction that unifies them has not yet been established.
\end{description}

\section{Toward Closed-Loop LLM Orchestration in the Cloud Continuum}
\label{sec:design-space}

Across the surveyed systems, the LLM plays the same role regardless of structural differences: 
it is restricted to interpretation, planning, and cross-layer reasoning, while resource 
control remains with RL policies
or with deterministic platform components. Three patterns recur. In a
\emph{semantic front-end}, the LLM compiles operator intent into a
machine-executable goal and then hands execution off (IntentContinuum, Habib et
al.). In a \emph{workflow pipeline}, several LLM stages produce a structured
deployment plan (SALLMA, AgentEdge). In a \emph{strategic orchestrator}, the LLM
issues subgoals or reward-shaping signals and revises them as agents report back
(AURA, CyberOps-Bots). Only this third pattern keeps the model coupled to a
delegated agent layer after the first directive, so it is the only one that can
support an agent-mediated control loop at the orchestration level (F2/F3). This is
why every F2/F3 system in our survey is an LLM2 orchestrator, while no front-end or
pipeline goes beyond F1.

The same observation accounts for the gap identified in
Section~\ref{sec:gap}: no LLM2 system with F2 or F3 feedback addresses
the Cloud Continuum. The three obstacles share a single cause: a
strategic orchestrator cannot yet operate across the Continuum because
nothing translates the heterogeneous per-tier signals into a unified
feedback abstraction the model can act on. AURA and CyberOps-Bots each
build such an abstraction within a single tier, but neither needs one
that spans the Continuum.

Our generic reference architecture in Fig.~\ref{fig:closed-loop} includes such a cross-tier feedback abstraction as a \emph{Mediator} component. As a layer of \emph{tier-specialised} RL agents is supervised by the orchestration layer, each agent trained on its own metrics and actions, one or more mediator instances collect the diverse outcomes.

The mediator extends the role of AURA's Centralised Alignment
Controller
from a single-tier feedback
provider to a cross-tier one: it ingests heterogeneous per-tier signals
and produces a structured representation that
pairs performance scores with the per-tier context the orchestrator needs
to revise its directives. Without this abstraction, incommensurable
per-tier signals would leave the orchestrator unable to determine which
tier or policy requires adjustment; the cross-tier mediator is therefore
the structural prerequisite for closing the loop across the Continuum.
Like the CAC in AURA, it is kept apart from the agents it evaluates, and that
separation places the channel at F3.

\section{Conclusion}
\label{sec:conclusion}
This paper addressed a relationship that the DRL literature on Continuum resource
management had left implicit: the architectural role an LLM assumes once it joins a
control loop that DRL and MARL agents already populate. We extended the taxonomy of
Wang et al.~\cite{wang2026drl} with two further dimensions capturing the AI Augmentation Paradigm
and the Feedback Channel, the latter stating whether and through
which component the execution feedback returns to the orchestrator. When applied to six
recent systems, the taxonomy isolated a specific gap. Closed-loop orchestration
exists only within a single resource tier, while intent-driven
Continuum management exists only as open-loop control or with platform-level feedback. 
Joining the two approaches in a multi-agent LLM orchestrator closing the loop
across IoT, edge, and cloud is still untried.

This
reflects a missing feedback abstraction rather than a settled negative result.
The components of closed-loop Continuum orchestration are individually
available, yet none of the surveyed systems combines them across the
heterogeneity that defines the Continuum. 
Two research targets follow. The primary --- closing the feedback loop in the LLM2-MA-Standard Cloud Continuum cell --- was formally characterized: a \emph{mediator} function is needed to reconcile heterogeneous signal sources towards the LLM. 
The secondary aim, which we regard as future work, is to extend the taxonomy to future LLM orchestrators using new solutions, e.g., an agent layer with distributed, federated or continual learning.

\begin{credits}
\subsubsection{\ackname}
This research was partially supported by the FIS2 Grant from the Italian Ministry of University and Research (MUR), Grant No. FIS2023-03382, under the project “Continual, Decentralized Compositionality for Sustainable Artificial Intelligence” with Prof. Vincenzo Lomonaco serving as Principal Investigator (PI).
\end{credits}

\bibliographystyle{splncs04}
\bibliography{bibliography}

@inproceedings{becattini2025sallma,
	title = {{SALLMA}: {A} {Software} {Architecture} for {LLM}-{Based} {Multi}-{Agent} {Systems}},
	shorttitle = {{SALLMA}},
	OPTurl = {https://ieeexplore.ieee.org/document/11029425},
	doi = {10.1109/SATrends66715.2025.00006},
	urldate = {2026-04-18},
	booktitle = {2025 {IEEE}/{ACM} {International} {Workshop} {New} {Trends} in {Software} {Architecture} ({SATrends})},
	author = {Becattini, Marco and Verdecchia, Roberto and Vicario, Enrico},
	month = apr,
	year = {2025},
	pages = {5--8},
}

@inproceedings{habib2025llmhrl,
	title = {{LLM}-{Based} {Intent} {Processing} and {Network} {Optimization} {Using} {Attention}-{Based} {Hierarchical} {Reinforcement} {Learning}},
	issn = {1558-2612},
	OPTdoi = {10.1109/WCNC61545.2025.10978505},
    note = {{DOI: 10.1109/WCNC61545.2025.10978505}, {ISSN: 1558-2612}},
    OPTnote = {{ISSN: 1558-2612}},
	urldate = {2026-04-18},
	booktitle = {2025 {IEEE} {Wireless} {Communications} and {Networking} {Conference} ({WCNC})},
	author = {Habib, Md Arafat and Iturria Rivera, Pedro Enrique and others},
    OPTauthor = {Ozcan, Yigit and Elsayed, Medhat and Bavand, Majid and Gaigalas, Raimundus and Erol-Kantarci, Melike},
	month = mar,
	year = {2025},
	pages = {1--6},
}

@article{gort2025agentedge,
	title = {{AgentEdge}: {Agentic} {AI} for {Service} {Orchestration} in the {Edge}-{Cloud} {Continuum}},
	volume = {2025},
	shorttitle = {{AgentEdge}},
	doi = {10.36227/techrxiv.175503000.00051702/v1},
	number = {0812},
	urldate = {2026-04-18},
	journal = {TechRxiv},
	publisher = {TechRxiv},
	author = {Gort, Berend and Kibalya, Godfrey M and Antonopoulos, Angelos},
	year = {2025},
}

@misc{peng2026cyberops,
	title = {Enhancing {Cloud} {Network} {Resilience} via a {Robust} {LLM}-{Empowered} {Multi}-{Agent} {Reinforcement} {Learning} {Framework}},
	url = {https://arxiv.org/abs/2601.07122v1},
	language = {en},
	urldate = {2026-04-22},
	journal = {arXiv.org},
	author = {Peng, Yixiao and Hu, Hao and Li, Feiyang and Cao, Xinye and Jiang, Yingchang and Tang, Jipeng and Nan, Guoshun and Liu, Yuling},
	month = jan,
	year = {2026},
}

@misc{wang2026drl,
	title = {Deep {Reinforcement} {Learning} for {Resource} {Management} in {IoT}-{Edge}-{Cloud} {Computing} {Continuum}: {A} {Taxonomy} and {Future} {Directions}},
	author = {Wang, Zhiyu and others},
	year = {2026},
	note = {under review},
    OPTurl = {https://clouds.cis.unimelb.edu.au/papers/DRL-RM-Continuum2026.pdf},
    url = {https://www.buyya.com/papers/DRL-RM-Continuum2026.pdf}
}

@inproceedings{akbari2025intentcontinuum,
	title = {{IntentContinuum}: {Using} {LLMs} to {Support} {Intent}-{Based} {Computing} {Across} the {Compute} {Continuum}},
	issn = {2836-3868},
	shorttitle = {{IntentContinuum}},
	doi = {10.1109/ICWS67624.2025.00079},
	urldate = {2026-04-18},
	booktitle = {2025 {IEEE} {International} {Conference} on {Web} {Services} ({ICWS})},
	author = {Akbari, Negin and Grundy, John and Cheema, Aamir and Toosi, Adel N.},
	month = jul,
	year = {2025},
	note = {{ISSN: 2836-3868}},
	pages = {573--583},
}

@misc{nourzad2025aura,
	title = {{AURA}: {Adaptive} {Unified} {Reasoning} and {Automation} with {LLM}-{Guided} {MARL} for {NextG} {Cellular} {Networks}},
	shorttitle = {{AURA}},
	doi = {10.48550/arXiv.2511.17506},
	urldate = {2026-04-09},
	publisher = {arXiv},
	author = {Nourzad, Narjes and Zong, Mingyu and Krishnamachari, Bhaskar},
	month = oct,
	year = {2025},
	OPTnote = {arXiv:2511.17506 [cs]},
}

@article{kephart2003vision,
  author  = {J. O. Kephart and D. M. Chess},
  title   = {The Vision of Autonomic Computing},
  journal = {IEEE Computer},
  volume  = {36},
  number  = {1},
  pages   = {41--50},
  year    = {2003},
  month   = jan
}

@misc{yao_react_2023,
	title = {{ReAct}: {Synergizing} {Reasoning} and {Acting} in {Language} {Models}},
	shorttitle = {{ReAct}},
	OPTdoi = {10.48550/arXiv.2210.03629},
	urldate = {2026-03-30},
	publisher = {arXiv},
	author = {Yao, Shunyu and Zhao, Jeffrey and others},
    OPTauthor = {Yu, Dian and Du, Nan and Shafran, Izhak and Narasimhan, Karthik and Cao, Yuan},
	month = mar,
	year = {2023},
	note = {{DOI: 10.48550/arXiv.2210.03629}},
    OPTnote = {arXiv:2210.03629 [cs]},
}

@inproceedings{sunehag2018vdn,
  author    = {Peter Sunehag and
               Guy Lever and others},
  OPTauthor =  {Audrunas Gruslys and
               Wojciech Marian Czarnecki and
               Vinicius Zambaldi and
               Max Jaderberg and
               Marc Lanctot and
               Nicolas Sonnerat and
               Joel Z. Leibo and
               Karl Tuyls and
               Thore Graepel},
  title     = {Value-Decomposition Networks For Cooperative Multi-Agent Learning
               Based On Team Reward},
  booktitle = {Proc. of the 17th Intl. Conf. on Autonomous
               Agents and MultiAgent Systems},
  OPTbooktitle = {Proc. of the 17th Intl. Conf. on Autonomous
               Agents and MultiAgent Systems ({AAMAS} 2018)},
  pages     = {2085--2087},
  publisher = {Intl. Foundation for Autonomous Agents and Multiagent Systems},
  year      = {2018},
  address   = {Stockholm, Sweden},
  month     = {July}
}

@inproceedings{rashid2018qmix,
  author    = {Tabish Rashid and
               Mikayel Samvelyan and others},
  OPTauthor = {
               Christian Schroeder de Witt and
               Gregory Farquhar and
               Jakob Foerster and
               Shimon Whiteson},
  title     = {{QMIX}: Monotonic Value Function Factorisation for Deep
               Multi-Agent Reinforcement Learning},
  booktitle = {Proc. of the 35th Intl. Conf. on Machine
               Learning ({ICML} 2018)},
  series    = {Proceedings of Machine Learning Research},
  ISSN      = {2640-3498},
  volume    = {80},
  pages     = {4295--4304},
  publisher = {PMLR},
  year      = {2018},
  address   = {Stockholm, Sweden},
  OPTurl       = {https://proceedings.mlr.press/v80/rashid18a.html}
}

@inproceedings{lowe2017maddpg,
  author    = {Ryan Lowe and
               Yi Wu and
               Aviv Tamar and others},
  optauthor  = {
               Jean Harb and
               Pieter Abbeel and
               Igor Mordatch},
  title     = {Multi-Agent Actor-Critic for Mixed Cooperative-Competitive
               Environments},
  booktitle = {Advances in Neural Information Processing Systems 30
               ({NeurIPS} 2017)},
  pages     = {6382--6393},
  year      = {2017},
  note      = {{ISBN: 9781510860964}},
  OPTurl       = {https://proceedings.neurips.cc/paper_files/paper/2017/hash/68a9750337a418a86fe06c1991a1d64c-Abstract.html}
}

@article{Lan2019ALBERTAL,
  title   = {ALBERT: A Lite BERT for Self-supervised Learning of Language Representations},
  author  = {Zhenzhong Lan and Mingda Chen and Sebastian Goodman and Kevin Gimpel and Piyush Sharma and Radu Soricut},
  journal = {CoRR},
  year    = {2019},
  volume  = {abs/1909.11942},
  url     = {https://arxiv.org/abs/1909.11942}
}

@article{chen2023frugalgpt,
  title={Frugalgpt: How to use large language models while reducing cost and improving performance},
  author={Chen, Lingjiao and Zaharia, Matei and Zou, James},
  journal={arXiv preprint arXiv:2305.05176},
  year={2023}
}

@inproceedings{NEURIPS2023_1b44b878,
 author = {Shinn, Noah and Cassano, Federico and others},
 OPTauthor = {Gopinath, Ashwin and Narasimhan, Karthik and Yao, Shunyu},
 booktitle = {Advances in Neural Information Processing Systems},
 OPTeditor = {A. Oh and T. Naumann and A. Globerson and K. Saenko and M. Hardt and S. Levine},
 pages = {8634--8652},
 publisher = {Curran Associates, Inc.},
 title = {Reflexion: language agents with verbal reinforcement learning},
 OPTurl = {https://proceedings.neurips.cc/paper_files/paper/2023/file/1b44b878bb782e6954cd888628510e90-Paper-Conference.pdf},
 volume = {36},
 year = {2023},
 note   = {{ISBN: 9781713899921}}
}

@article{ullah_orchestration_2023-1,
	title = {Orchestration in the {Cloud}-to-{Things} compute continuum: taxonomy, survey and future directions},
	volume = {12},
	issn = {2192-113X},
	OPTurl = {https://doi.org/10.1186/s13677-023-00516-5},
	doi = {10.1186/s13677-023-00516-5},
	number = {1},
	journal = {Journal of Cloud Computing},
	author = {Ullah, Amjad and Kiss, Tamas and Kovács, József and Tusa, Francesco and Deslauriers, James and Dagdeviren, Huseyin and Arjun, Resmi and Hamzeh, Hamed},
	month = sep,
	year = {2023},
	pages = {135},
}

\end{document}